\documentclass[12pt,english]{article}
\pdfoutput=1
\usepackage[T1]{fontenc}
\usepackage[latin9]{inputenc}
\usepackage[letterpaper]{geometry}
\usepackage{url}
\usepackage{bm}
\usepackage{graphicx}
\usepackage{setspace}

\makeatletter
\usepackage{times}
\usepackage{graphics,natbib}

\newcommand{\matern}{Mat\'{e}rn }
\newcommand{\maternp}{Mat\'{e}rn}

\usepackage{babel}

\makeatother

\usepackage{babel}
\begin{document}

\title{Spatial models for point and areal data using Markov random fields
on a fine grid}

\author{Christopher J. Paciorek}

\date{Department of Statistics, University of California, Berkeley}

\maketitle

\begin{abstract}
I consider the use of Markov random fields (MRFs) on a fine grid to
represent latent spatial processes when modeling point-level and areal
data, including situations with spatial misalignment. Point observations
are related to the grid cell in which they reside, while areal observations
are related to the (approximate) integral over the latent process
within the area of interest. I review several approaches to specifying
the neighborhood structure for constructing the MRF precision matrix,
presenting results comparing these MRF representations analytically,
in simulations, and in two examples. The results provide practical
guidance for choosing a spatial process representation and highlight
the importance of this choice. In particular, the results demonstrate
that, and explain why, standard CAR models can behave strangely for
point-level data. They show that various neighborhood weighting approaches
based on higher-order neighbors that have been suggested for MRF models
do not produce smooth fields, which raises doubts about their utility.
Finally, they indicate that an MRF that approximates a thin plate spline
compares favorably to standard CAR models and to kriging under many
circumstances. 
\end{abstract}

\section{Introduction}

Markov random field (MRF) models (also called conditional autoregressive
(CAR) models) are the dominant approach to analyzing areally-aggregated
spatial data, such as disease counts in administrative units. In contrast,
point-referenced data are generally modeled using Gaussian process
(GP) models that posit a continuous latent underlying spatial surface,
of which the popular approach of kriging for spatial prediction is
one variant. 

GP models are computationally challenging for large datasets because
of manipulations involving large covariance matrices, and there has
been a large body of recent work attempting to reduce the computational
burden through reduced rank approximations \citep{Kamm:Wand:2003,Bane:etal:2008,Sang:Huan:2012},
inducing sparseness in the covariance (covariance tapering) \citep{Furr:etal:2006,Kauf:etal:2008,Sang:Huan:2012},
and approximate likelihood approaches \citep{Stei:etal:2004}, among
others. In contrast, MRF models work with the precision matrix directly,
so calculation of the likelihood is computationally simple. In MCMC
implementations, one can exploit the Markovian structure to sample
the value of the field for each area sequentially or exploit the sparsity
of the precision matrix when sampling the field values jointly \citep{Rue:Held:2005}.

Given the computational attractiveness of the MRF approach, it is
appealing to consider its use for point-referenced data as well. One
issue lies in how to define the neighborhood structure for a set of
point-referenced observations, but a second more fundamental issue
lies in the fact that the model structure changes when one changes
the number of sites under consideration. The solution that I highlight
here is to relate the point-referenced observations to an underlying
regularly-gridded surface modeled as an MRF.

Considering an underlying gridded surface for areally-aggregated data
is also appealing. \citet{Kels:Wake:2002} argue for the use of a
smooth underlying surface to model areal data, with each areal observation
related to the average of the surface over the defined area. They
found approximations to the integrals involved and worked with an
underlying GP representation; related work includes \citet{Fuen:Raft:2005}
and \citet{Hund:etal:2012}. Here I relate areal observations to an
underlying MRF on a fine grid, approximating the necessary integrals
as a simple weighted average of the MRF values for the grid cells
that overlap each area. Such an approach also allows one to model
multiple areally-aggregated datasets with different boundary structures.
In summary, the approach of using an MRF on a fine grid provides an
aggregation-consistent model for point or areal data (or a mix of
both) and is a strategy also suggested in \citet{Besa:Mond:2005}.

\begin{figure}
\includegraphics[scale=0.6]{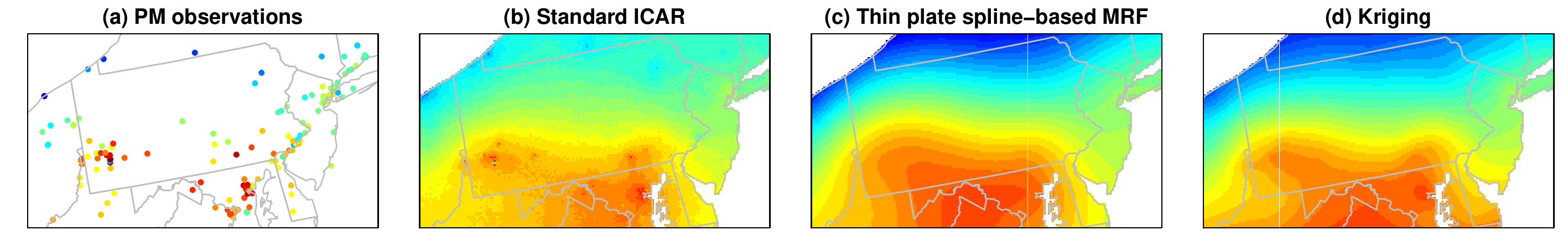}

\caption{Example fits to particulate matter (PM) air pollution point observations
(a) using the standard first-order MRF on a fine regular grid (b),
an MRF approximation to a thin plate spline, again represented on
a fine regular grid (c), and using kriging with an exponential covariance
(d). \label{fig:Example-of-fitting}}
\end{figure}

The most common form of MRF represents the spatial dependence structure
such that areas that share a boundary are considered neighbors, with
an area conditionally independent (given its neighbors) of any non-neighboring
areas, a so-called first-order neighborhood structure. Fig.~\ref{fig:Example-of-fitting}
shows the results of using this MRF structure on a fine regular grid
to model point-referenced data. The fitted smooth surface is not visually
appealing, in contrast to an MRF that approximates a thin plate spline
\citep{Rue:Held:2005} and to kriging. The results are consistent
with \citet{Besa:Mond:2005}, who show that the intrinsic (i.e., improper)
first-order MRF on a two-dimensional regular grid produces spatial
fields whose distribution approaches two-dimensional Brownian motion
(the de Wijs process) asymptotically as the grid resolution increases.
Given this continuous but non-differentiable representation of the
underlying surface, the local heterogeneity of the surface estimate
in Fig.~\ref{fig:Example-of-fitting} is not surprising. However,
note that \citet{Besa:Mond:2005} and Besag in his comments on \citet{Digg:etal:2010}
argue that the de Wijs process is preferable to GPs in the \matern
class. 

A common alternative to this standard nearest-neighbor structure is
to extend the neighborhood structure beyond bordering areas \citep{Pett:etal:2002,Hraf:Cres:2003,Song:etal:2008}.
Such higher-order neighborhood structure might be expected to produce
more smooth process representations, but I show that straightforward
higher-order neighborhoods do not achieve this. An alternative that
I highlight here is an MRF approximation to a thin plate spline (TPS)
that involves only nearby grid cells as neighbors \citep{Rue:Held:2005,Yue:Spec:2010}.
Finally, \citet{Lind:etal:2011} have recently developed a powerful
theory and methodology for approximating GPs in the \matern class
that will likely be widely used. The thin plate spline approximation
is a limiting case of the \citet{Lind:etal:2011} representation,
and my results shed light on the distinctions between standard first-order
MRF models, the thin plate spline approximation, and GP representations.

In this paper, I present a general model for both areal and point-referenced
data that deals simply with spatial misalignment by using an MRF on
a fine regular grid (Section 2.1). In the literature, first-order
MRF representations are widely-used for areal data, generally with
little consideration of the properties of the latent process or resulting
suitability for a given application. In light of this and of the lack
of smoothness of the popular first-order MRF seen in Fig.~\ref{fig:Example-of-fitting},
I compare several existing approaches for the MRF neighborhood structure
(Section 2.2) using both analytic calculations and simulations (Section
3) in the context of the general model presented here. The results
suggest that the thin plate spline approximation often performs well
and help to explain the strange behavior of the first-order MRF. Importantly
for situations in which a smooth process representation is desired,
higher-order neighborhood structures can have unappealing properties
when they are not constructed as approximations to particular spatial
surface representations. Finally, MRF models perform as well as kriging
under a variety of circumstances for both point and areal data. I
close by discussing computation (Section 4) and presenting two examples,
one with point data and one with areal data (Section 5). Online supplementary material \citep{Paci:2013} contains the R code for all analyses and figures as well as the data for the pollution data only, as the breast cancer data are not publicly available.

\section{Spatial model for point and areal data}

\subsection{Model structure}

Here I present a basic model for exponential family data. Let $\mu_{i}=E(Y_{i}|\bm{X}_{i},\bm{g})$
be related via a link function, $h(\cdot)$, to a linear predictor:
\begin{equation}
h(\mu_{i})=\bm{X}_{i}^{\top}\bm{\beta}+\bm{K}_{i}\bm{g},\label{eq:GLMM1}
\end{equation}
where $\bm{K}_{i}$ is the $i$th row of a mapping matrix, $\bm{K}$,
discussed further below. I represent the unknown, latent spatial process,
$g(\cdot)$, as a piecewise constant surface on a fine rectangular
grid,
\begin{equation}
\bm{g}\sim\mathcal{N}(\bm{0},(\kappa\bm{Q})^{-})\label{eq:GLMM2}
\end{equation}
where $\bm{Q}$ is an MRF precision matrix and $\kappa$ a precision
parameter, recognizing the potential singularity of $\bm{Q}$ by using
the generalized inverse notation. Computational issues related to
the singularity are discussed in Section \ref{sec:Computation}.

For a point-referenced datum, $\bm{K}_{i}$ will be a sparse vector
with a single 1 that matches the location of the observation to the
grid cell in which it falls. Note that one may include covariates
in $\bm{X}$ that can help to account for within grid cell heterogeneity.
For an areally-aggregated datum, I consider the relevant functional
of the surface to be the average value of the underlying surface over
the area $A_{i}$: $g^{*}(A_{i})\equiv\int_{A_{i}}g(\bm{s})d\bm{s}$.
Using the piecewise representation, this can be approximated as $g^{*}(A_{i})\approx\sum_{j\in A_{i}}w_{j}g_{j}$
where $j\in A_{i}$ indexes the grid cells that the $A_{i}$ overlaps
and $w_{j}$ is the proportion of $A_{i}$ that falls in the $j$th
grid cell. Hence the non-zero elements of $\bm{K}_{i}$ contain the
proportions, $\{w_{j}\}$, as the appropriate elements. If desired,
one could interpolate between the values of $\bm{g}$ at the grid
cell centroids for a more accurate approximation.

\subsection{Potential MRF models}

Here I present the MRF models that I consider for $\bm{g}$ and their
corresponding precision matrices, $\bm{Q}$. I consider only intrinsic
models with singular precision matrices. These specify improper priors
with respect to one or more linear combinations of the process values,
as the eigenvectors of $\bm{Q}$ with zero eigenvalues have infinite
prior variance \citep{Bane:etal:2003}. My focus on intrinsic models
is motivated by noting that proper first-order MRF models tend not
to allow high correlations between neighbors unless the precision
matrices are close to singular \citep{Bane:etal:2003,Wall:2004}.
Furthermore, intrinsic models represent the conditional mean for the
process in a given area as a weighted average of the process values
of the neighboring areas with weights summing one, while proper models
have weights summing to less than one.

\begin{enumerate}
\item Traditional intrinsic CAR model (ICAR): The most commonly-used MRF
model is a simple first-order model that treats any two areas that
share a border as neighbors. The corresponding precision matrix has
diagonal elements, $Q_{ii}$, equal to the number of neighbors for
the $i$th area, while $Q_{ij}=-1$ (the negative of a weight of one)
when areas $i$ and $j$ are neighbors and $Q_{ij}=0$ when they are
not. On a grid, the simplest version of this model treats the four
nearest grid cells as neighbors (i.e., cardinal neighbors). \citet{Besa:Mond:2005}
show that the model is asymptotically equivalent to two-dimensional
Brownian motion, and \citet{Lind:etal:2011} show that this model
approximates a GP in the \matern class (\ref{eq:Matern}) with the
spatial range parameter $\rho\to\infty$ and differentiability parameter
$\nu\to0$. 
\item Extended neighborhood model (DICAR and HICAR): One might generalize
the first-order Markovian structure of the ICAR model to allow for
higher-order dependence by considering areas that do not share a border
but are close in some sense to be neighbors. At its simplest, this
simply introduces additional values of $-1$ off the diagonal of $\bm{Q}$
as in \citet{Song:etal:2008}. I will call this model the higher-order
ICAR (HICAR). A more nuanced version would have the weight for a pair
of areas depend on the distance between the two areas (usually declining
with distance), such as Euclidean distance or the number of intervening
cells between the two areas \citep{Pett:etal:2002,Hraf:Cres:2003}.
Then $Q_{ij}=-\delta(i,j)$ where $\delta(\cdot,\cdot)$ is the chosen
weight or distance function. I will term this model the distance-based
ICAR (DICAR) and implement the model using the function in \citet{Hraf:Cres:2003},
$\delta(i,j)=-d_{ij}^{\log.05/\log r}$, where $d_{ij}$ is the distance
between area centroids and $r$ is the (user-chosen) maximum distance
at which the weight is non-zero.

\item Thin plate spline approximation (TPS-MRF): \citet[Sec.~3.4.2]{Rue:Held:2005}
present a second-order model where the weights on neighbors of different
orders are constructed so the model approximates a thin plate spline.
The approach considers a discretized approximation of the penalty
function used in deriving the thin plate spline solution. This penalty
function is an integral of second partial derivatives of the unknown
surface, so the discretization produces a prior precision matrix based
on second order difference operators. In the discretized approach,
the nearest cardinal neighbors have a weight of $8$ ($Q_{ij}=-8)$,
the nearest diagonal neighbors a weight of $-2$ ($Q_{ij}=2$) and
the second nearest cardinal neighbors a weight of $-1$ ($Q_{ij}=1)$.
Note the presence of negative weights, unlike in most MRF models with
higher order neighborhood structures. \citet[App. C]{Paci:Liu:2012}
describe the derivation of the full $\bm{Q}$ matrix, including boundary
effects, in detail. In one dimension, this model corresponds to the
widely-used second-order auto-regressive model (an IID model on second
differences) \citep{Bres:Clay:1993}. \citet{Lind:etal:2011} show
that this model approximates a GP in the \matern class (\ref{eq:Matern})
with $\nu=1$ and the spatial range parameter $\rho\to\infty$.
\end{enumerate}

In this work I compare MRF models to GPs in the \matern class, where
I parameterize the \matern correlation function as 
\begin{equation}
R(d)=\frac{1}{\Gamma(\nu)2^{\nu-1}}\left(\frac{2\sqrt{\nu}d}{\rho}\right)^{\nu}\mathcal{K}_{\nu}\left(\frac{2\sqrt{\nu}d}{\rho}\right),\label{eq:Matern}
\end{equation}
where $d$ is Euclidean distance, $\rho$ is the spatial range parameter,
and $\mathcal{K}_{\nu}(\cdot)$ is the modified Bessel function of
the second kind, whose order is the smoothness (differentiability)
parameter, $\nu>0$. $\nu=0.5$ gives the exponential covariance.
Note that the \citet{Lind:etal:2011} approach provides a computational
strategy for fitting an approximate GP model via a MRF representation
but in the material that follows I consider the explicit GP rather
than an MRF approximation to it.

\section{Comparing MRF Structures}

This section compares the different MRF models using a variety of tools, starting with an assessment of the implied eigenstructure and smoothing kernels of the models (Section 3.1). The comparison sheds light on the smoothing behavior of the ICAR model and shows that the extended neighborhood models have unappealing properties. In Section 3.2 I compare predictive performance of the models, both under an oracle setting with the optimal choice of smoothing parameter and in simulations in which all parameters are estimated.

\subsection{Covariance and smoothing properties}

\subsubsection{Eigenstructure}

To better understand the spatial dependence structures implied by
the various MRF precision matrices, I quantify the magnitude of the
variability for different scales (frequencies) of spatial variability
by considering the eigendecomposition of $\bm{Q}$. Given the model
$\bm{g}\sim\mathcal{N}(\bm{0},\bm{Q}^{-})$, we can consider the eigendecomposition,
$\bm{Q}^{-}=\bm{\Gamma}\bm{\Lambda}^{-}\bm{\Gamma}^{\top}$. To generate
realizations of $\bm{g}$, we have $\bm{g}=\bm{\Gamma}\bm{u}$ for
$\bm{u}\sim\mathcal{N}(\bm{0},\bm{\Lambda}^{-})$. Thus the inverse
eigenvalues quantify the magnitude of variability associated with
patterns or modes of variability represented by the eigenvectors. 

Empirical exploration indicates that the inverse eigenvalues decline
as the frequency of variability represented in the eigenvectors increases.
Therefore to visualize how different MRF models weight variability
at different frequencies, I plot the ordered inverse eigenvalues.
Because the matrix $\bm{Q}$ is multiplied by a scalar precision,
only the relative magnitudes are of interest, so I scale the inverse
eigenvalues such that the 100th largest inverse eigenvalue for each
precision matrix is taken to be equal to one. While the eigenvectors
for the various precision matrices are not the same, empirical exploration
shows that they are quite similar and represent very similar spatial
scales of variation for a given position in the ordering of the eigenvector/value
pairs. Furthermore, projection onto a common set of eigenvectors (from
the ICAR model) confirms that the slight differences in eigenvectors
between models do not impact the results shown next.

In the comparisons, I compare the TPS-MRF model to a GP with \matern
correlation with $\nu=2$, rather than $\nu=1$ as would be natural
given the \citet{Lind:etal:2011} relationship. I do this because
thin plate splines represent smooth functions and the \matern has
$M$ mean square derivatives when $\nu>M$ \citep[p. 32]{Stei:1999}.
I compare the ICAR model to a GP with an exponential correlation (\matern
with $\nu=0.5$) despite the \citet{Lind:etal:2011} result relating
the ICAR model to a \matern covariance with $\nu\to0$ because the
\matern covariance is only valid for $\nu>0$. 

\begin{figure}
\includegraphics[scale=0.75]{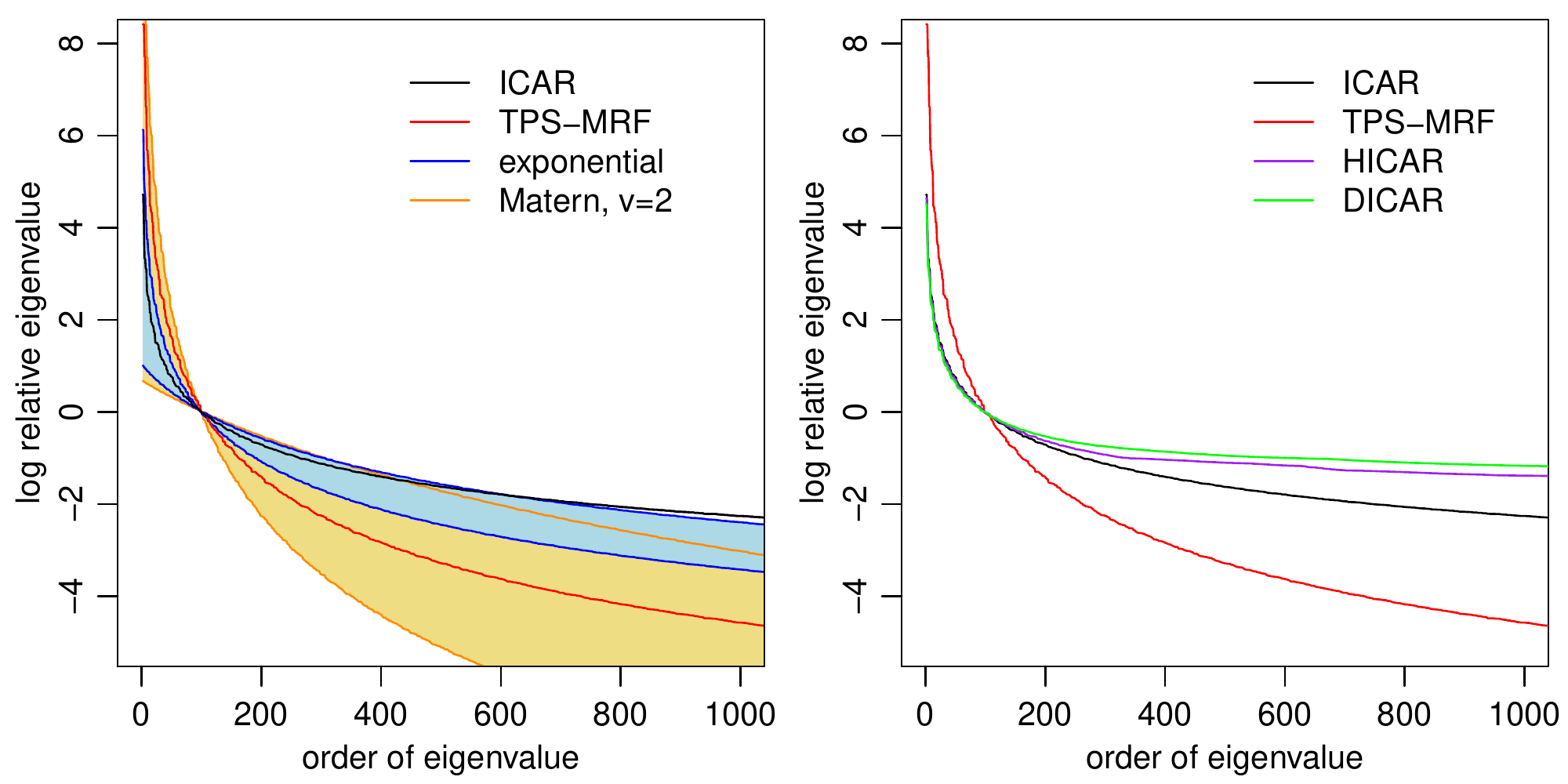}

\caption{(a) Log inverse eigenvalues (relative to the 100th largest inverse
eigenvalue) for the ICAR and TPS-MRF models, in comparison with the
log inverse eigenvalues for the \matern covariance (with $\nu=2$)
and exponential covariance models. The yellow-orange shading, bounded
by orange lines, indicates the range of eigenvalue curves obtained
for the \matern as the range parameter, $\rho$, varies from the
size of the full domain in one dimension to 1/25 of the domain in
one dimension. Similarly, the blue shading, bounded by dark blue lines,
is for the exponential covariance as the range varies from the full
domain to 1/25 of the domain. (b) Log inverse eigenvalues (relative
to the 100th largest inverse eigenvalue) for the ICAR and TPS-MRF
models, both repeated from (a);  the HICAR model with grid cells
up to and including three units in distance considered neighbors;
and the DICAR model with non-zero weights that decay with distance
up to and including five units in distance. Only the first 1000 inverse
eigenvalues are plotted, as by the 1000th, the features represented
in the eigenvectors occur within groups of 5-10 grid cells, near the
limit of scales that could be resolved in a gridded representation.
\label{fig:eigenval}}
\end{figure}

Fig.~\ref{fig:eigenval} plots the size of the ordered inverse eigenvalues
for different MRF precision matrices and in comparison to GP models
for a regular spatial grid of dimension $75\times75$. The TPS-MRF
puts more weight on the lower frequency eigenvectors and less weight
on the higher frequency eigenvectors than the ICAR. This is not surprising
given the relationship of the ICAR model to Brownian motion and the
smoothness of splines. The TPS-MRF eigenvalue curve lies within the
set of eigenvalue curves (for varying values of the range parameter,
$\rho$) from the \matern model with $\nu=2$. At low frequency,
the ICAR eigenvalue curve lies within the set of eigenvalue curves
from the exponential model, while the ICAR model puts more weight
on high frequency eigenvectors than the exponential model, consistent
with the ICAR approximating a \matern covariance with $\nu\to0$
\citep{Lind:etal:2011}. 

Particularly interesting is the behavior of the higher-order and distance-based
ICAR variations, which behave similarly to the ICAR model (Fig.~\ref{fig:eigenval}b).
At low frequency, the curves coincide, while the HICAR and DICAR models
put more weight on the higher frequencies than the ICAR model. This
indicates that one cannot use these approaches to represent surfaces
smoother than those represented by the ICAR model. In fact, based
on this analysis, it is unclear why one would use these representations,
given that the motivation for their use has been to provide more smoothness
than the ICAR model.

\subsubsection{Equivalent kernels\label{sub:Equivalent-kernels}}

To understand the smoothing behavior of the various models, I next
consider their equivalent kernels, which quantify the local averaging
that the models do to make predictions. Under a normal likelihood
and ignoring covariates for simplicity, $\bm{Y}\sim\mathcal{N}(\bm{K}\bm{g},\tau^{2}\bm{I})$.
When there are observations at all the grid cells, the smoothing matrix,
$\bm{S}$, in $\hat{\bm{g}}=\bm{S}\bm{y}$ can be expressed as 
\begin{eqnarray*}
\bm{S} & = & \frac{1}{\tau^{2}}\left(\kappa\bm{Q}+\frac{1}{\tau^{2}}\bm{I}\right)^{-1}=(\lambda\bm{Q}+\bm{I})^{-1},
\end{eqnarray*}
where $\lambda\equiv\tau^{2}\kappa$. For the GP models, we have $\hat{\bm{g}}=\hat{\mu}\bm{1}+\sigma^{2}\bm{R}_{\theta}(\sigma^{2}\bm{R}_{\theta}+\tau^{2}\bm{I})^{-1}(\bm{y}-\hat{\mu}\bm{1})$
so 
\begin{eqnarray*}
\bm{S} & = & \sigma^{2}\bm{R}_{\theta}(\sigma^{2}\bm{R}_{\theta}+\tau^{2}\bm{I})^{-1}=(\lambda\bm{R}_{\theta}^{-1}+\bm{I})^{-1},
\end{eqnarray*}
where $\tau^{2}$ is the error variance, $\sigma^{2}$ is the marginal
variance of the GP, and $\bm{R}_{\theta}$ is the correlation matrix,
a function of parameter(s), $\theta$. $\lambda\equiv\tau^{2}/\sigma^{2}$
can be thought of as a smoothing parameter, as in the MRF model. Note
that the sum of the weights is not one when expressing $\bm{S}$ as
above, because it ignores $\hat{\mu}$, which is also linear in the
observations. I ignore this component of the prediction as it involves
adding and subtracting a constant that does not vary by location.

\begin{figure}
\includegraphics[scale=0.8]{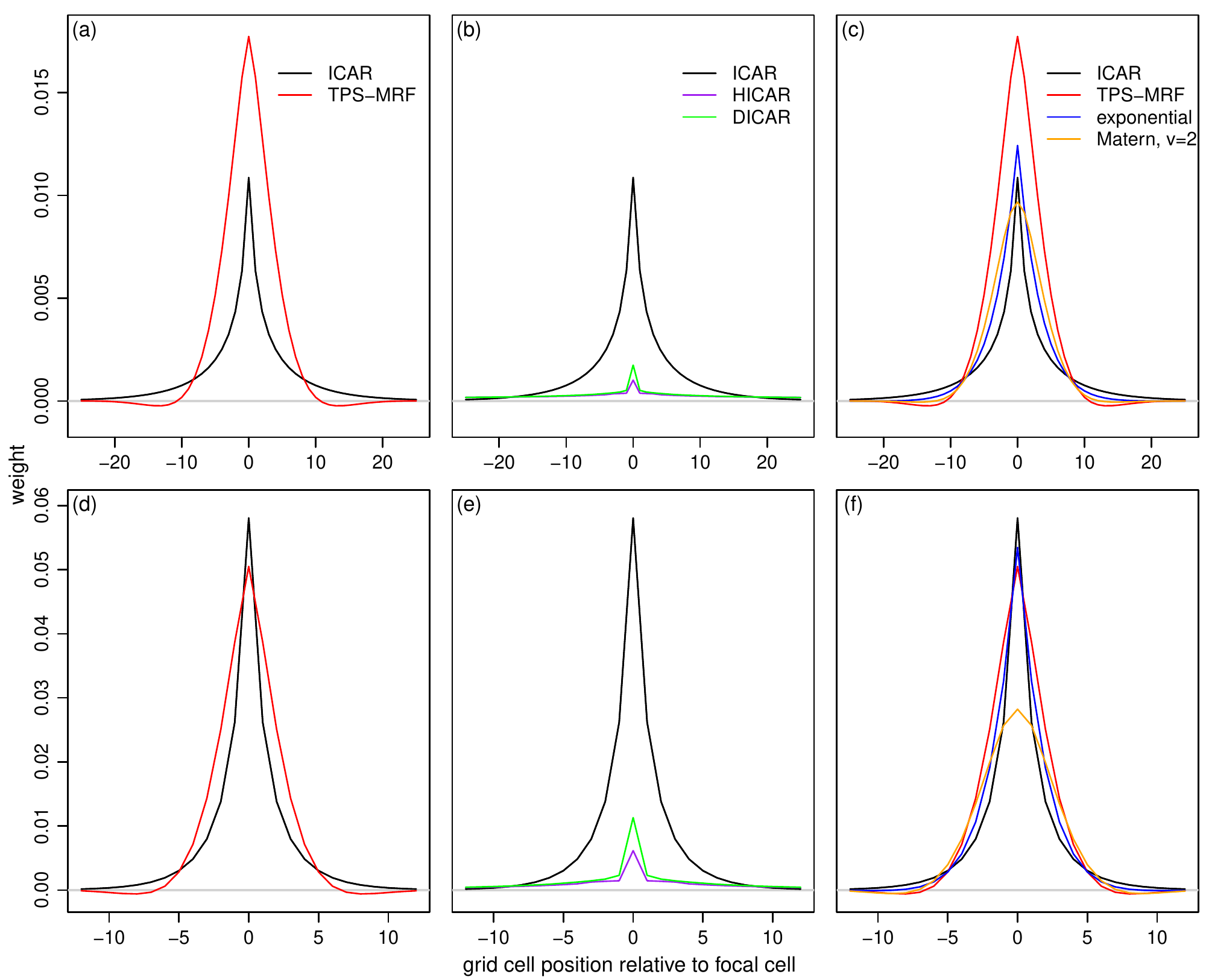}

\caption{Equivalent kernel cross-sections in one of the cardinal directions
for various models with (a-c) $\lambda=\exp(4)$. which produces less
localized weighting and (d-f) $\lambda=\exp(2)$, which produces more
localized weighting. In (b) and (e), the HICAR model has grid cells
up to and including three units in distance considered neighbors,
and the DICAR model has decaying non-zero weights up to and including
five units in distance. In (c) and (f), the \matern (with $\nu=2$)
and exponential covariance models have range, $\rho$, set to one-tenth
of the domain size in one dimension. \label{fig:equivKernels}}
\end{figure}

Fig.~\ref{fig:equivKernels} shows the various kernels, plotted as
a function of one of the cardinal directions, with the other direction
held fixed, for two values of the smoothing parameter, $\lambda$.
Fig.~\ref{fig:equivKernels}a,d shows that the ICAR kernel puts less
weight near the focal cell and more further from the focal cell, compared
to the TPS-MRF, but with a spike at the focal cell. This behavior
helps to explain the bulls-eye effect and the greater shrinkage towards
an overall mean in the gaps between observations seen for the ICAR
model in Fig.~\ref{fig:Example-of-fitting}. The TPS-MRF kernel shows
some small-magnitude negative weights, which is consistent with the
negative weights in the equivalent kernels for spline smoothing and
Gaussian process smoothing \citep{Silv:1984,Soll:Will:2005}. In
Fig.~\ref{fig:equivKernels}b,e, we see that the HICAR (with neighbors
within three units) and DICAR (with non-zero weights within five units)
models place very little weight near the focal observation and spread
their weight very widely. The result is that their kernels are even
more extreme than the ICAR kernel in making predictions that heavily
weight observations far from the focal location. Results are similar
but not as extreme for HICAR and DICAR models with smaller neighborhoods.
As with the eigenvector analysis, this suggests the HICAR and DICAR
models have little practical appeal. Fig.~\ref{fig:equivKernels}c,f
compares the ICAR and TPS-MRF to the equivalent kernels for GP models
with $\rho$ set to one-tenth of the size of the domain in one dimension. The
ICAR model shows some similarity to the exponential-based GP in terms
of tail behavior and the spike at the focal cell. The TPS-MRF equivalent
kernels are rather different than the \maternp-based GP model with
$\nu=2$, but more similar in terms of tail behavior than when compared
to the exponential-based model.

\begin{figure}
\includegraphics[scale=0.6]{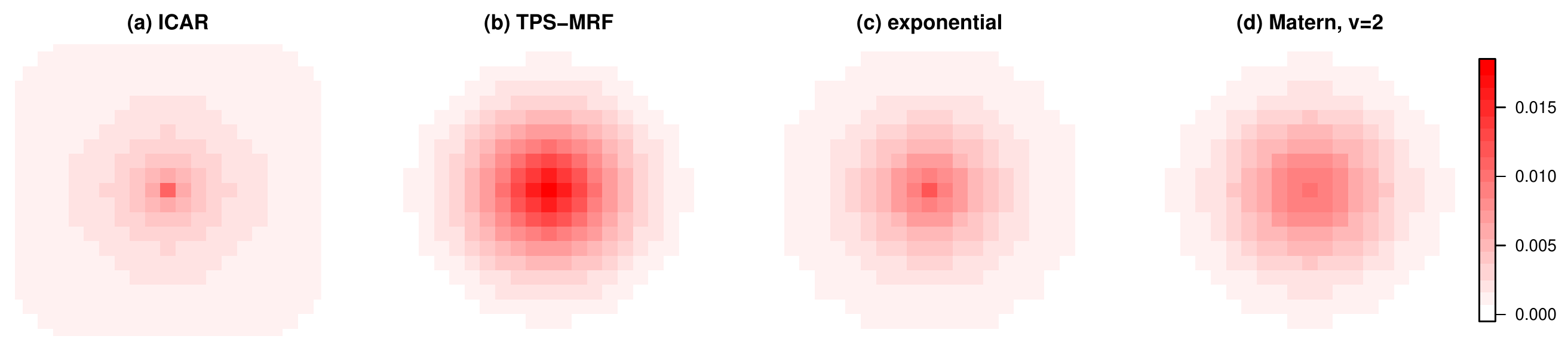}

\caption{Equivalent kernels in two dimensions for $\lambda=\exp(4)$: (a) ICAR,
(b) TPS-MRF, (c) exponential covariance with range, $\rho$, equal
to one-tenth of the domain in one dimension, and (d) \matern covariance
with $\nu=2$ and that same range.\label{fig:equivKernels2}}
\end{figure}

Fig.~\ref{fig:equivKernels2} reinforces these points, showing image
plots of the equivalent kernels in two dimensions for the ICAR and
TPS-MRF models, as well as the \matern and exponential covariance
models. The ICAR kernel puts little weight near the focal cell but
spreads positive weight further from the focal cell than the other
approaches. The exponential is qualitatively similar, but less extreme
than the ICAR. The TPS-MRF and \matern models are somewhat similar,
but the TPS-MRF puts more weight near the focal cells.

\subsection{Assessment of predictive ability\label{sec:oracleAnal}}

Here I compare predictive performance of the ICAR and TPS-MRF models
with GP models, via analytic calculation of squared error (using an
oracle estimator of the smoothing parameter) and in simulations in
which all parameters are estimated.

\subsubsection{Data-generating scenarios}

I consider the simple generative model for $\bm{Y}=\{Y_{1},\ldots,Y_{n}\}$:

\begin{eqnarray}
\bm{Y} & \sim & \mathcal{N}(\bm{g},\tau^{2}\bm{I})\nonumber \\
\bm{g} & \sim & \mathcal{N}(\bm{X\beta},\bm{C}),\label{eq:generative}
\end{eqnarray}
with $\bm{C}=\sigma^{2}\bm{R}_{\theta}$, where $\bm{R}_{\theta}$
is a correlation matrix based on the \matern correlation function
with $\nu\in\{0.5,\,2\}$ ($\nu=0.5$ is equivalent to the exponential
correlation function). Here $\bm{X}$ represents spatial basis functions,
including an overall mean and possibly linear and higher-order functions
of the spatial coordinates, as in ordinary kriging and universal kriging.
Using the unit square as the domain, I consider five scenarios: $n=100$
with locations uniformly sampled, $n=100$ with observations sampled
according to a Poisson cluster process (PCP) with parent intensity
of 25 and standard normal kernels with standard deviation of 0.05,
$n=1000$ with uniform and PCP sampling, and $n=100$ areal observations
on a coarse $10\times10$ grid overlaid on a regular $100\times100$
grid. For each scenario, I use a full factorial design with respect
to $\nu\in\{0.5,2\}$, $\rho\in\{0.005,0.02,0.08,0.32,1.28,2.56\}$,
and $\tau^{2}\in\{0.05^{2},0.15^{2},0.45^{2},1.35^{2}\}$. For $n=100$
and $n=1000$, I use $10$ replicates to average over the random sampling
of locations; the resulting Monte Carlo standard errors are small
relative to the point estimates of the reported SSE values. Data are
simulated over the unit square in a continuous fashion, while predictions
are considered at the grid cell centroids. For simplicity, I simulate
with $\sigma^{2}=1$ and $\bm{\beta}=\bm{0}$.

\subsubsection{Model fitting}

I consider fitting the data using either the ICAR or TPS-MRF models
on a regular $100\times100$ grid:
\begin{eqnarray*}
\bm{Y} & \sim & \mathcal{N}(\bm{K}\bm{g},\tau^{2}\bm{I}),\\
\bm{g} & \sim & \mathcal{N}(\bm{X\beta},(\kappa\bm{Q})^{-})
\end{eqnarray*}
where $\bm{K}$ is a mapping matrix that relates observation locations/areas
to grid cells. $\bm{X\beta}$ represents the overall mean in the ICAR
model and the mean plus linear functions of the spatial coordinates
in the TPS-MRF model. One can express the best prediction for $\bm{g}$
as

\begin{eqnarray}
\hat{\bm{g}} & = & E(\bm{g}|\bm{Y},\cdot)=(\bm{K}^{\top}\bm{K}+\lambda\bm{Q})^{-1}(\bm{K}^{\top}\bm{Y}+\lambda\bm{QX}\hat{\bm{\beta}})\nonumber \\
 & = & (\bm{K}^{\top}\bm{K}+\lambda\bm{Q})^{-1}(\bm{K}^{\top}+\lambda\bm{QX}((\bm{KX})^{\top}\bm{\Sigma}^{*-1}\bm{KX})^{-1}(\bm{KX})^{\top}\bm{\Sigma}^{*-1})\bm{Y}\nonumber \\
 & = & (\bm{K}^{\top}\bm{K}+\lambda\bm{Q})^{-1}\bm{K}^{\top}\bm{Y}\nonumber \\
 & \equiv & \bm{S}_{\lambda}\bm{Y}.\label{eq:ghat}
\end{eqnarray}
where $\bm{\Sigma}^{*}\equiv\bm{I}+\bm{K}(\lambda\bm{Q})^{-}\bm{K}^{\top}$,
$\lambda\equiv\tau^{2}\kappa$, and the terms involving $\bm{X}$
drop out, with estimation of $\bm{\beta}$ absorbed into $\bm{g}$,
thereby avoiding non-identifiability. 

The expected sum of squared errors, averaging over randomness in observations
and randomness in latent spatial process realizations, can be expressed
in terms of a single parameter, $\lambda$, that needs to be estimated,
plus the parameters of the data-generating model:

\begin{eqnarray}
\mbox{SSE}(\lambda) & = & E_{g}E_{Y}((\hat{\bm{g}}-\bm{g})^{\top}(\hat{\bm{g}}-\bm{g}))\nonumber \\
 & = & E_{g}E_{Y}(\bm{Y}^{\top}\bm{S}_{\lambda}^{\top}\bm{S}_{\lambda}\bm{Y}-2\bm{g}^{\top}\bm{S}_{\lambda}\bm{Y}+\bm{g}^{\top}\bm{g})\nonumber \\
 & = & E_{g}(\tau^{2}\bm{S}_{\lambda}^{\top}\bm{S}_{\lambda}-2\bm{g}^{\top}\bm{S}_{\lambda}\bm{Kg}+\bm{g}^{\top}\bm{g}+\bm{g}^{\top}\bm{K}^{\top}\bm{S}_{\lambda}^{\top}\bm{S}_{\lambda}\bm{Kg})\nonumber \\
 & = & \tau^{2}\mbox{tr}(\bm{S}_{\lambda}^{\top}\bm{S}_{\lambda})-2\mbox{tr}(\bm{S}_{\lambda}\bm{KC})+\mbox{tr}(\bm{C})+\mbox{tr}(\bm{K}^{\top}\bm{S}_{\lambda}^{\top}\bm{S}_{\lambda}\bm{KC})\nonumber \\
 &  & -2(\bm{X\beta})^{\top}\bm{S}_{\lambda}\bm{KX\beta}+(\bm{X\beta})^{\top}\bm{X\beta}+(\bm{X\beta})^{\top}\bm{K}^{\top}\bm{S}_{\lambda}^{\top}\bm{S}_{\lambda}\bm{KX\beta}\nonumber \\
 & = & \tau^{2}\mbox{tr}(\bm{S}_{\lambda}^{\top}\bm{S}_{\lambda})-2\mbox{tr}(\bm{S}_{\lambda}\bm{KC})+\mbox{tr}(\bm{C})+\mbox{tr}(\bm{K}^{\top}\bm{S}_{\lambda}^{\top}\bm{S}_{\lambda}\bm{KC}).\label{eq:mse}
\end{eqnarray}
In the derivation, I make use of the fact that $\bm{S}_{\lambda}\bm{KX}=\bm{X}$
(because $\bm{QX}=\bm{0}$ when the columns of $\bm{X}$ are in the
space spanned by the eigenvectors of $\bm{Q}$). To simplify the presentation
of results, I consider an oracle result based on optimizing the SSE
(\ref{eq:mse}) across all possible values of $\lambda$ for each
of the ICAR and TPS-MRF models, thereby using the best overall $\lambda$
for a given generative setting. In this, I assume that $\tau^{2}$
and $\sigma^{2}$ are known.

Analogous calculations for SSE can be done when fitting the model
using the GP approach (thereby fitting with the same model used to
generate observations). For the GP approach, I consider predictions
at the grid centroids but take the data locations to be the actual
locations. As a result the expected SSE can be expressed in similar
form to (\ref{eq:mse}), but with $\bm{S}$ involving $\bm{R}_{\theta}^{-1}$
in place of $\bm{Q}$, $\lambda\equiv\frac{\tau^{2}}{\sigma^{2}}$,
and a slight bit of additional complexity to account for the correlation
between the process values at the data locations and those at the
grid cell centroids. Since the GP is the generating model, it provides
a baseline for comparison with the MRF models, so I also assume that
$\lambda=\frac{\tau^{2}}{\sigma^{2}}$ is known, as are $\bm{\theta}=\{\rho,\nu\}$
and $\bm{\beta}=\bm{0}$. 

The oracle assessment just described presumes good choices of the
penalty parameter for each representation and does not assess the
effect of hyperparameter estimation on performance. Therefore, I also
present basic simulation results under the data-generating scenarios
described above, with the exception that for the areal scenario, I
do not fit the GP model because of computational constraints. I use
$100$ simulations; the resulting Monte Carlo standard errors are
small relative to the reported SSE results. The unknowns are $\bm{g}$,
$\tau^{2}$, $\bm{\beta}$ (for the GP model), and $\lambda=\kappa\tau^{2}$
(for the MRF) or $\lambda=\tau^{2}/\sigma^{2}$ (for a GP model).
Integrating over $\bm{g}$ to obtain a marginal likelihood in terms
of $\tau^{2}$ and $\lambda$ and profiling over $\tau^{2}$ (and
$\bm{\beta}$ for the GP) gives a likelihood that can be numerically
maximized with respect to $\lambda$. One can then use (\ref{eq:ghat})
to estimate $\bm{g}$ (with the analogous quantity for the GP case).
In fitting the GP model, I use the true $\nu$.

\subsubsection{Results for point observations \label{sub:Results-for-point}}

\begin{figure}
\includegraphics[scale=0.8]{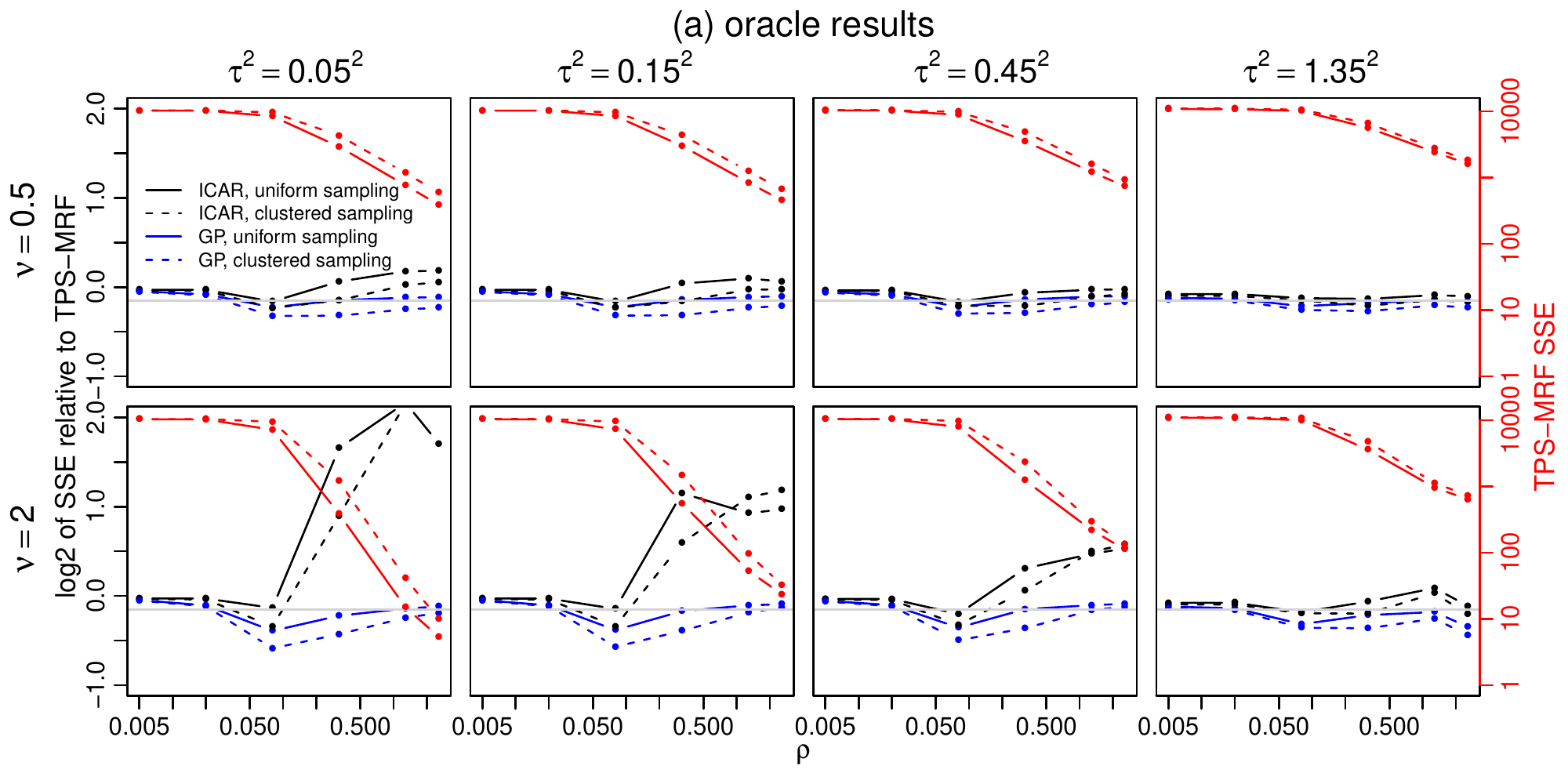}

\includegraphics[scale=0.8]{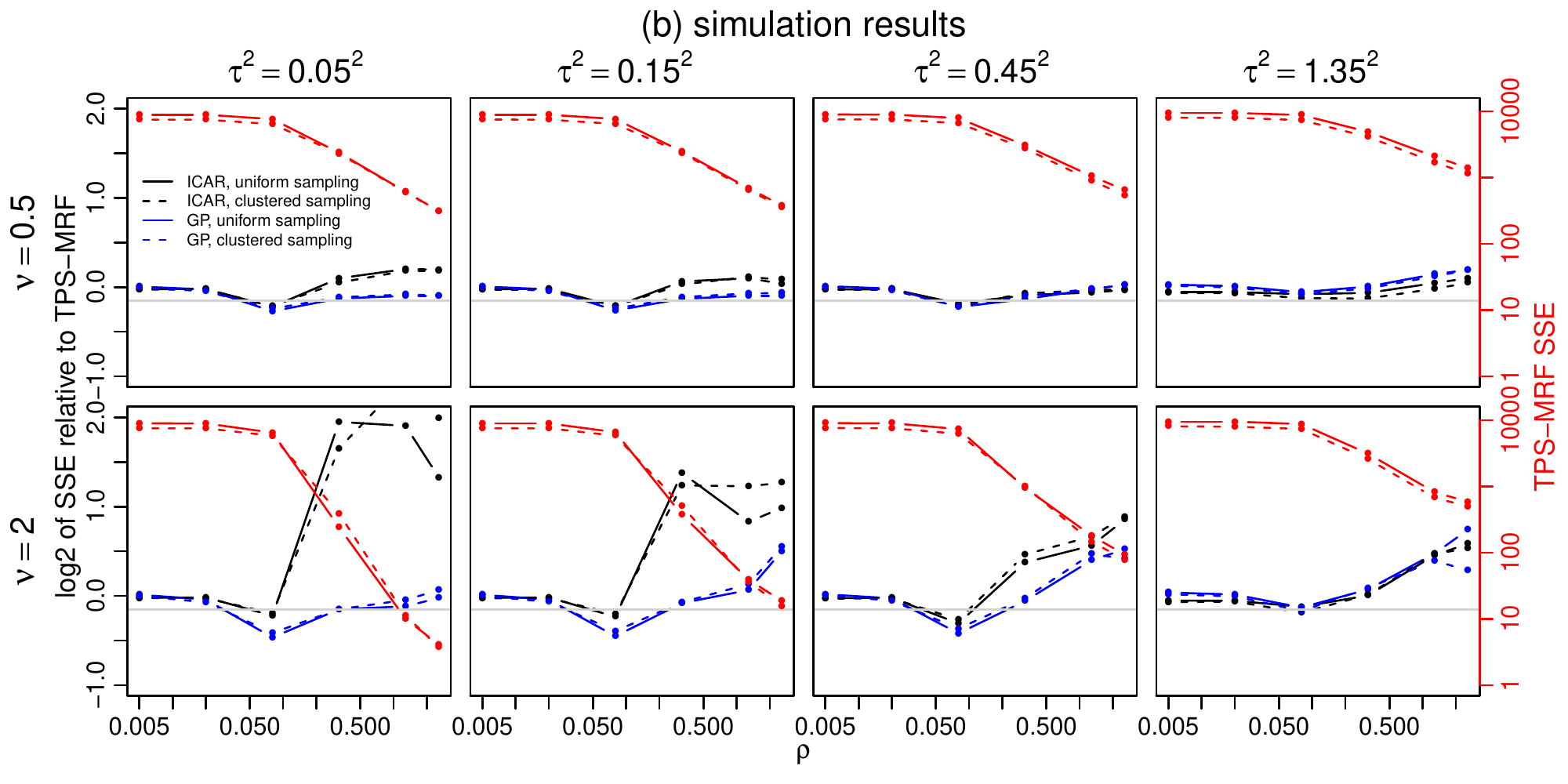}

\caption{Predictive performance for $n=100$ point observations: (a) oracle
results and (b) simulation results. Plots show the log (base 2) of
the ratio of SSE for the ICAR (black) and GP (blue) models relative
to the TPS-MRF and absolute SSE for the TPS-MRF for reference (in
red, with axis labels on the right side). In each subplot, $\nu$
varies with the row and $\tau^{2}$ with the column. The horizontal
grey line corresponds to the SSE being 90\% of the SSE of the TPS-MRF
as an informal cutoff below which other models perform substantially
better. The SSE in (b) is computed for locations within the convex
hull of the observations in a given simulation. \label{fig:Predictive-performance-for-100}}
\end{figure}

\begin{figure}
\includegraphics[scale=0.8]{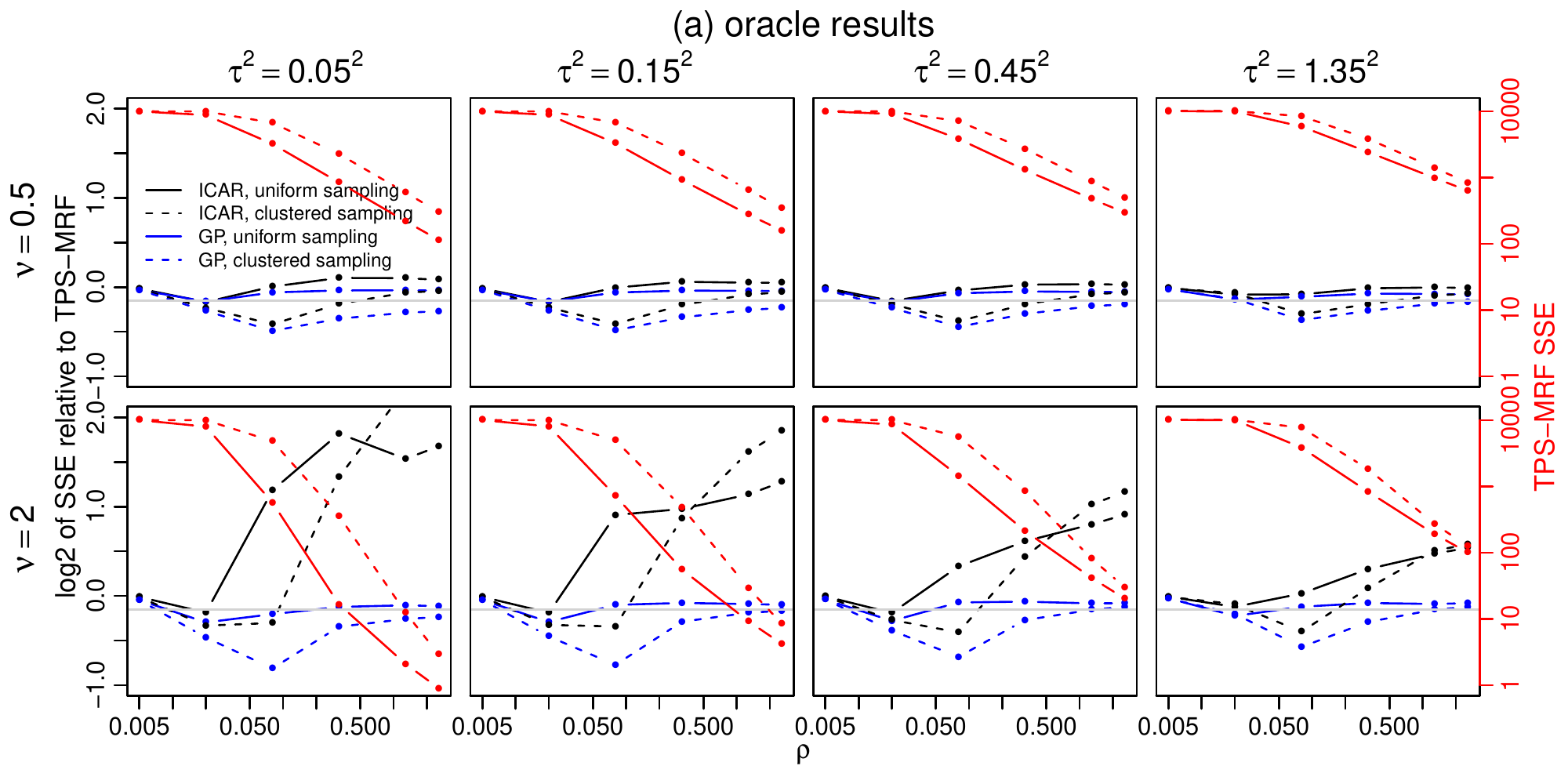}

\includegraphics[scale=0.8]{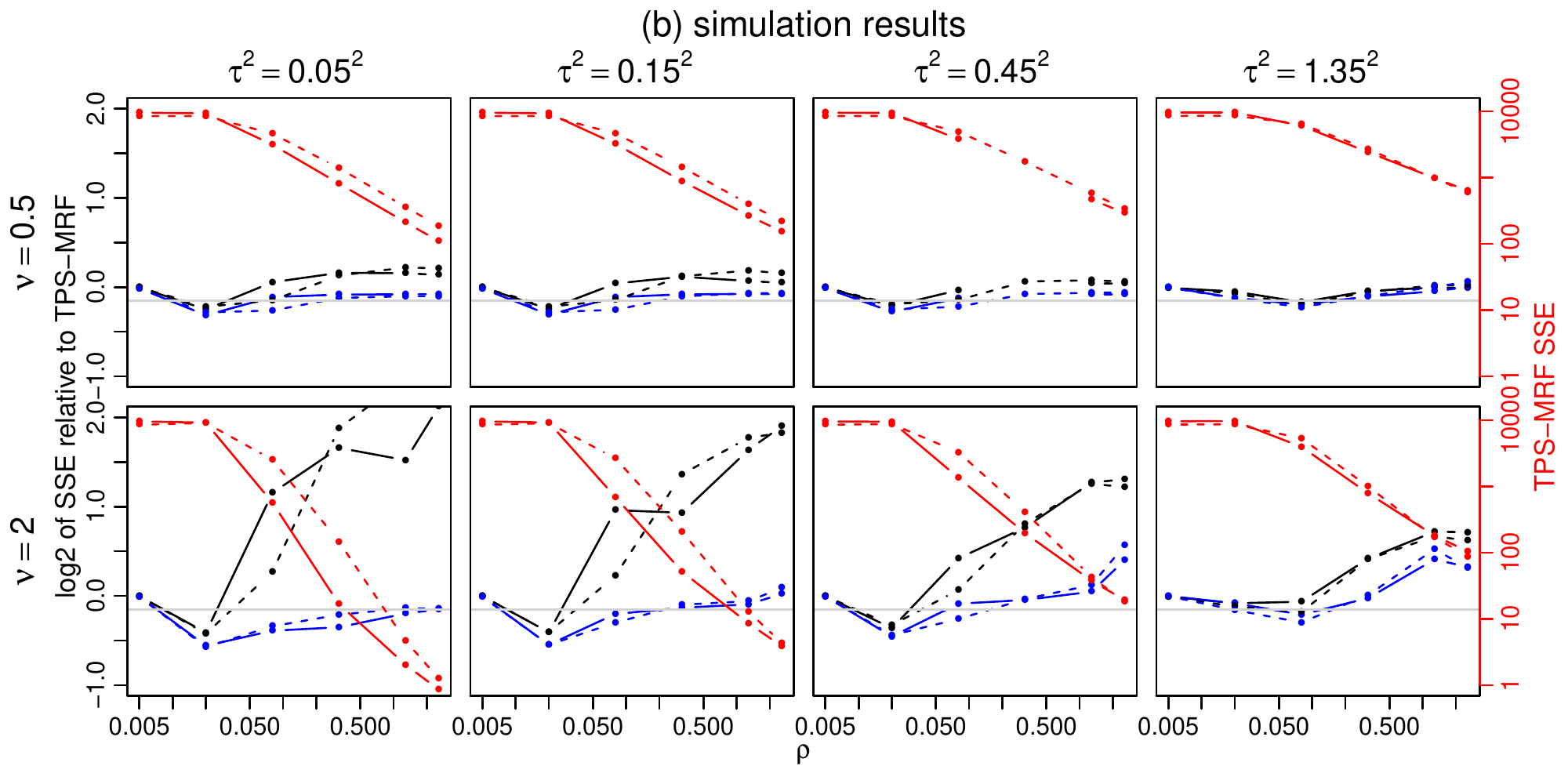}

\caption{Predictive performance for $n=1000$ point observations: (a) oracle
results and (b) simulation results. Details are as in Fig.~\ref{fig:Predictive-performance-for-100}.
\label{fig:Predictive-performance-for-1000}}
\end{figure}

For $n=100$ and $n=1000$, respectively, Figs.~\ref{fig:Predictive-performance-for-100}-\ref{fig:Predictive-performance-for-1000}
show the ratio of the SSE using the ICAR to that using the TPS-MRF
as well as the ratio of the SSE using the true GP model to that using
the TPS-MRF. We see that with uniformly distributed locations, in
general the TPS-MRF either matches the SSE of the ICAR or improves
upon it. The TPS-MRF strongly outperforms the ICAR when $\nu=2$,
the range is moderate to large, and the noise variance is not too
large. The one case in which the ICAR generally beats the TPS-MRF
is for $\rho=0.08$ for $n=100$ and $\rho=0.02$ for $n=1000$, particularly
for smaller values of $\tau^{2}$ and $\nu=2$, but note that in these
cases the absolute SSE (red lines) is close to the SSE of an intercept-only
null model for both models. For $\nu=0.5$, which produces locally
heterogeneous surfaces for which we would expect the ICAR model to
perform well, we see that the two MRF models perform fairly similarly.
Clustering appears to decrease the ratio of SSE of the ICAR relative
to the TPS-MRF. This is likely a result of the fact that the TPS-MRF
approximates a spline, with no constraint on the spline fit in any
large gaps with no observations. Unlike a Gaussian process, a spline
does not revert to an overall estimate of the mean when it is far
from any observations, which may lead to poor interpolation. Comparing
the TPS-MRF to the GP, the GP generally performs better, which is
not surprising given that it is the generative model and is fit based
on the true hyperparameter values, but particularly for uniformly-sampled
locations, the TPS-MRF is competitive. For larger values of both $\rho$
and $\tau^{2}$, with $\nu=2$, we see that the TPS-MRF actually outperforms
the GP in the simulations, despite the GP being the generative model,
suggesting that for smooth surfaces there is a cost to having to estimate
$\rho$ (in particular the GP tends to underestimate $\rho$ and therefore
undersmooth), as indicated by the fact that using a GP with known
parameter values for these same simulations results in the GP outperforming
the TPS-MRF (not shown). 

\begin{figure}
\includegraphics[scale=0.75]{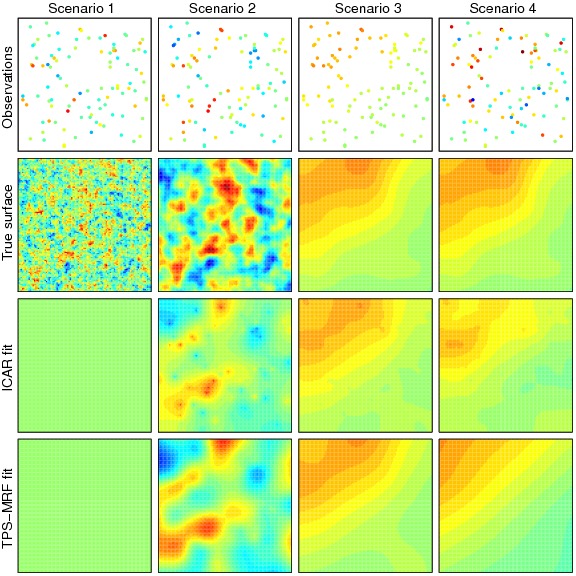}

\caption{Example ICAR and TPS-MRF fits (rows) for four generative scenarios (columns) with uniform
sampling of locations: (1) $\nu=0.5,\,\rho=0.02,\,\tau^{2}=0.15^{2}$,
(2) $\nu=2,\,\rho=0.08,\,\tau^{2}=0.15^{2}$, (3) $\nu=2,\,\rho=1.28,\,\tau^{2}=0.15^{2}$,
(4) $\nu=2,\,\rho=1.28,\,\tau^{2}=1.35^{2}$. \label{fig:Example-fits}}
\end{figure}

I now interpret these results further in the context of some example
fits. For small values of the range, the oracle fit is a constant
surface, and both MRF models do this and perform similarly, with SSE
approximately equal to $E_{g}\bm{g}^{\top}\bm{g}$, the SSE of $\hat{\bm{g}}=\bm{0}$;
see the example fits in Fig.~\ref{fig:Example-fits}, first column.
Neither model has any predictive ability, and SSE is essentially squared
bias. For slightly larger range values, the ICAR model does modestly
better (example fits in Fig.~\ref{fig:Example-fits}, second column),
with the TPS-MRF oversmoothing (and experiencing some boundary issues)
and the more local behavior and mean reversion of the ICAR model (recall
the results in Section~\ref{sub:Equivalent-kernels})
being beneficial. Then, as the range increases the TPS improves upon
the ICAR, most notably when the true surface is smooth ($\nu=2$)
(Fig.~\ref{fig:Example-fits}, third column), with the ICAR showing
the bulls-eyes seen in Fig.~\ref{fig:Example-of-fitting}. If the
noise variance is also very large, then the advantage of the TPS when
$\nu=2$ moderates, with the TPS oversmoothing (Fig.~\ref{fig:Example-fits},
fourth column). When $\nu=0.5$, performance of the ICAR and TPS-MRF
are comparable because the TPS oversmooths, but follows the larger-scale
patterns better than the ICAR, while the ICAR more closely follows
the local variability in the vicinity of the observations (not shown).

In terms of uncertainty characterization, in the simulations both
the ICAR and TPS-MRF give the nominal 95\% coverage for prediction
intervals (not shown). However, coverage for the true function, $\bm{g}$,
is quite low in some situations (not shown). In particular, coverage
is well below the nominal coverage levels for both the ICAR and TPS-MRF
for $\rho\leq0.02$, the settings in which both models estimate a
constant surface. The TPS-MRF also shows undercoverage for larger
values of $\rho$ when $\nu=0.5$. The good predictive coverage but
poor function coverage of the TPS-MRF occurs because the model assumes
a smooth underlying surface and attributes some of the true variability
in the function surface to the error component, thereby giving standard
errors for the function values that are too small. \citet{Bane:etal:2010}
noted a similar problem for reduced rank kriging, with an inflation
in the estimated nugget variance. This suggests that one interpret
the TPS-MRF uncertainty as relating to the larger-scale spatial variability
and accept that one is not able to characterize uncertainty about
finer-scale variation.

\subsubsection{Results for areal observations\label{sub:Results-for-areal}}

\begin{figure}
\includegraphics[scale=0.8]{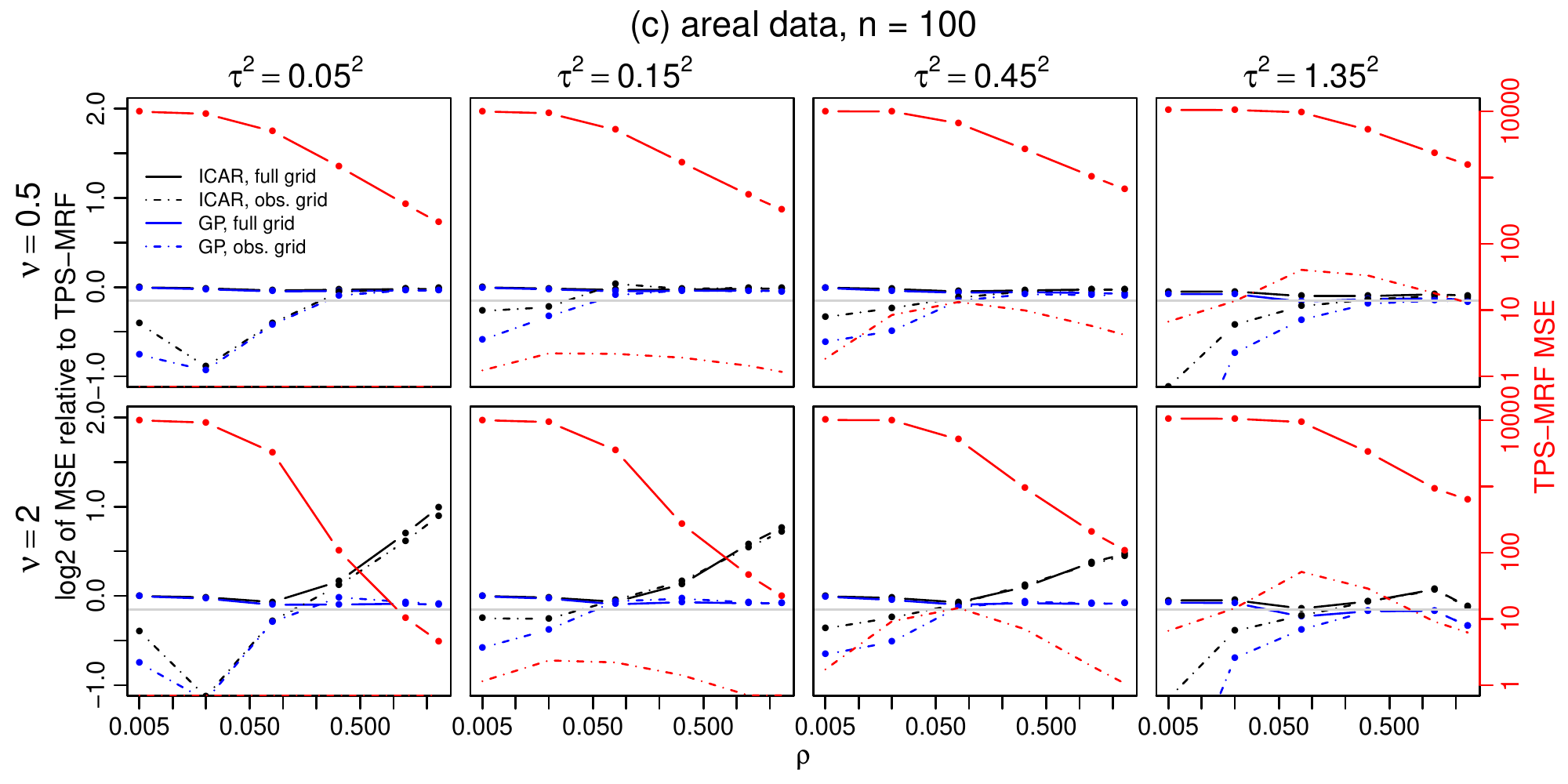}

\includegraphics[scale=0.8]{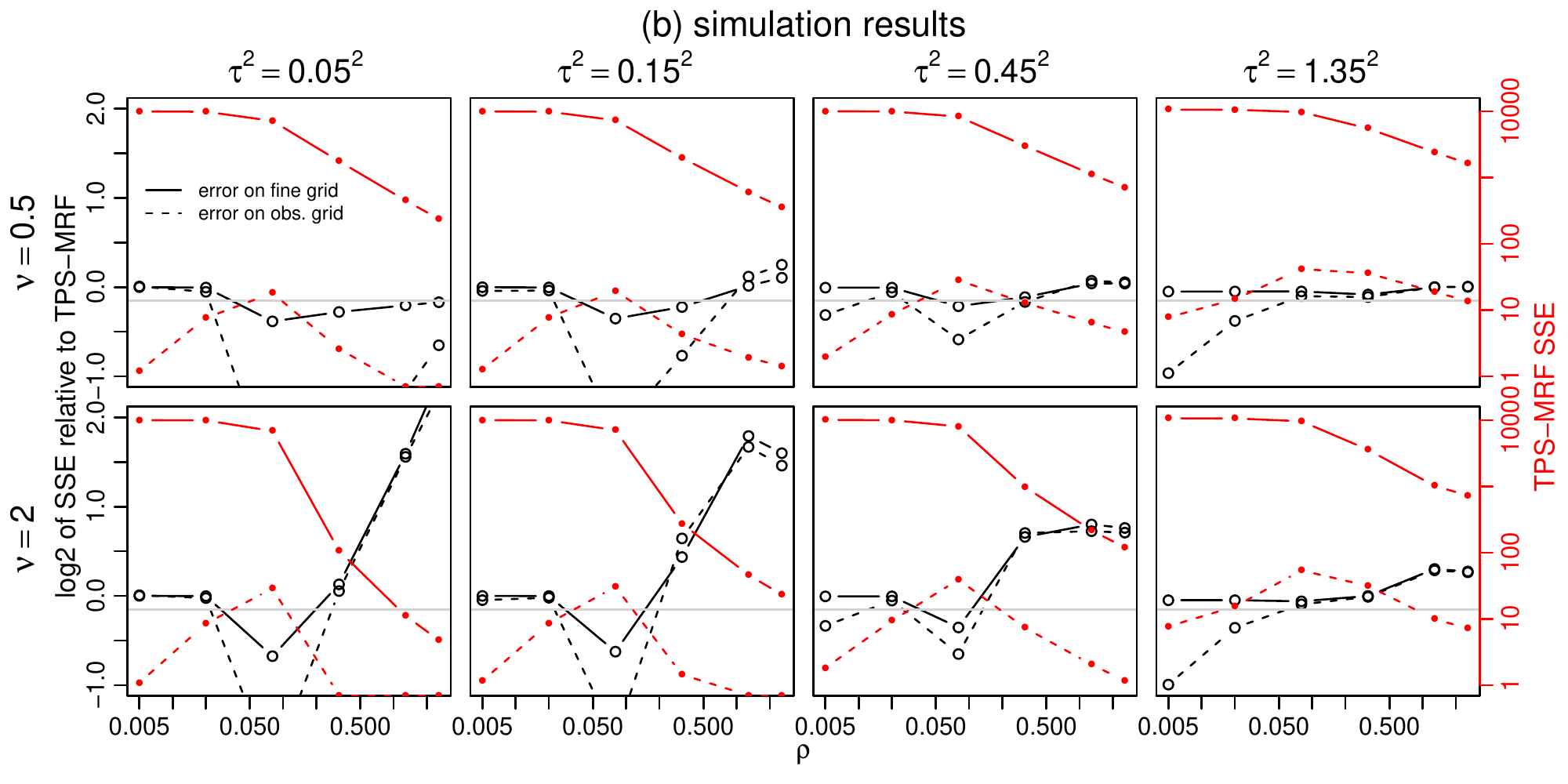}

\caption{Predictive performance for areal data: (a) oracle results and (b)
simulation results. Plots show the log (base 2) of the ratio of SSE
for the ICAR (black) and (for point observations only) GP (blue) models
relative to the TPS-MRF and absolute SSE for the TPS-MRF for reference
(in red, with axis labels on the right side). Solid lines show SSE
for latent process at the cell centroids of the fine grid, while dashed
lines show SSE for the values of the latent process averaged within
each grid cell of the coarse grid. Additional details are as in Fig.~\ref{fig:Predictive-performance-for-100}.\label{fig:areal}}
\end{figure}

Finally, for areal data, the TPS-MRF model outperforms the ICAR model
in some cases, primarily for larger values of $\rho$, while the ICAR
outperforms the TPS-MRF at $\rho=0.08$ in the simulations but not
the oracle results (Fig.~\ref{fig:areal}). The relative advantages
and disadvantages of the two models are more pronounced in the simulations,
suggesting the difficulty of parameter estimation in the areal context.
The primary setting in which the TPS-MRF performs poorly is at $\rho=0.08$,
particularly when the noise level is low. In these settings, at the
aggregated resolution of the observations the true surface is fairly
heterogeneous between neighboring areas and the ICAR follows the data
more closely, while the TPS-MRF oversmooths. However, note that in
these settings, the SSE is large, indicating that neither model is
able to fit the data well.

Coverage of the TPS-MRF prediction intervals is at the nominal 95\%
level (with the exception of one setting), but for large values of
$\rho$ and smaller values of $\tau^{2}$, the ICAR model often follows
the data exactly, estimating $\tau^{2}=0$ and giving very low coverage
(not shown). Considering coverage for the true $\bm{g}$, results
are similar to the situation for point observations for both models,
with nominal coverage maintained only for larger values of $\rho$,
with the ICAR model showing somewhat more settings that maintain nominal
coverage (not shown).

\section{Computational considerations\label{sec:Computation}}

\subsection{Normal data\label{sub:Normal-data}}

Consider an MRF model for $\bm{g}$, $\bm{g}\sim\mathcal{N}_{m}(\bm{0},(\kappa\bm{Q})^{-})$,
where the zero mean is justified by (\ref{eq:ghat}). If the observations
are normally distributed, $\bm{Y}\sim\mathcal{N}_{n}(\bm{X\beta}+\bm{Kg},\tau^{2}\bm{I})$,
we have conjugacy and can integrate $\bm{g}$ out of the model to
obtain a marginal likelihood with which to do maximization or MCMC
on the hyperparameter space. Note that one could also use the INLA
methodology to quickly approximate the posterior without MCMC \citep{Rue:etal:2009},
but here I explore the computations needed for maximum likelihood
and for 'exact' inference via MCMC.

The marginal precision for $\bm{Y}$ can be expressed, based on the
Sherman-Morrison-Woodbury formula, as 
\[
\bm{\Sigma}_{\lambda}^{-1}=\frac{1}{\tau^{2}}\left(\bm{I}-\bm{K}(\lambda\bm{Q}+\bm{K}^{\top}\bm{K})^{-1}\bm{K}^{\top}\right).
\]
 For maximization, we can express the maxima for $\bm{\beta}$ and
$\tau^{2}$ as functions of $\bm{\Sigma}_{\lambda}^{-1}$ and therefore
of $\lambda$,
\begin{eqnarray*}
\hat{\bm{\beta}}_{\lambda} & = & (\bm{X}^{\top}\bm{\Sigma}_{\lambda}^{-1}\bm{X})^{-1}\bm{X}^{\top}\bm{\Sigma}_{\lambda}^{-1}\bm{Y}\\
\hat{\tau}_{\lambda}^{2} & = & \frac{(\bm{Y}-\bm{X}\hat{\bm{\beta}}_{\lambda})^{\top}\bm{\Sigma}_{\lambda}^{-1}(\bm{Y}-\bm{X}\hat{\bm{\beta}}_{\lambda})}{n-c},
\end{eqnarray*}
where $c$ is the number of zero eigenvalues of $\bm{Q}$. Both of
these quantities can be calculated efficiently based on a sparse Cholesky
decomposition of $\lambda\bm{Q}+\bm{K}^{\top}\bm{K}$ in the expression
for $\bm{\Sigma}_{\lambda}^{-1}$, because $\bm{K}^{\top}\bm{K}$
will generally be sparse. If all the data are point locations, $\bm{K}^{\top}\bm{K}$
is diagonal, with the diagonal entries counting the number of observations
falling in each grid cell. If all the data are areal observations,
only off-diagonal elements of $\bm{K}^{\top}\bm{K}$ corresponding
to pairs of grid cells that are overlapped by a common areal observation
are non-zero. 

The marginal profile likelihood as a function of $\lambda$ alone
is proportional to 
\[
\frac{\lambda^{(m-c)/2}}{(\hat{\tau}_{\lambda}^{2})^{(n-c)/2}|\lambda\bm{Q}+\bm{K}^{\top}\bm{K}|^{1/2}}
\]
where the determinant can be calculated efficiently based on the already-computed
sparse Cholesky decomposition.

MCMC calculations rely on similar quantities that can be computed
efficiently. To draw from the posterior of $\bm{g}$ off-line, given
$\kappa$, $\tau^{2}$, and $\bm{\beta}$, we have 
\[
\bm{g}\sim\mathcal{N}((\bm{K}^{\top}\bm{K}+\kappa\tau^{2}\bm{Q})^{-1}\bm{K}^{\top}(\bm{Y}-\bm{X\beta}),\tau^{2}(\bm{K}^{\top}\bm{K}+\lambda\bm{Q})^{-1}),
\]
which can be done efficiently based on the same Cholesky decomposition
as above.

The computational limitation in the proposed MRF approach is the ability
to work with a sparse matrix $\bm{K}^{\top}\bm{K}+\lambda\bm{Q}$
whose size scales with the number of grid cells. Thus the approach
is limited only by the resolution at which one wishes to do prediction
rather than by the sample size, which is similar to the computational
constraint in reduced rank kriging, where the computational cost scales
with the number of knots. For small $n$ and large $m$, it may be
more computationally efficient to represent the data covariance as
\[
\bm{\Sigma}=\tau^{2}\left(\bm{I}+\frac{1}{\lambda}\bm{K}\bm{Q}^{-}\bm{K}^{\top}\right)
\]
and precompute the $n$ by $n$ matrix $\bm{K}\bm{Q}^{-}\bm{K}^{\top}$,
while in the iterations of an MCMC or optimization computing the Cholesky
of the dense matrix $\bm{\Sigma}$. Note that here we need to use
the generalized inverse, thereby assigning zero variance to the eigenvectors
of $\bm{Q}$ corresponding to zero eigenvalues and hence necessitating
inclusion of the relevant terms in the mean of $\bm{g}$.

\subsection{Non-normal data}

The general model (\ref{eq:GLMM1}-\ref{eq:GLMM2}) is a GLMM, where
$\bm{K}$ and $\bm{g}$ play the roles of the random effects design
matrix and random effects, respectively. In this case the covariance
of the random effects has spatial structure and is specified in terms
of a precision matrix. Both likelihood-based and Bayesian inference
for GLMMs is computationally challenging because inference involves
a high-dimensional integral with respect to the random effects that
cannot be expressed in closed form. 

While the PQL approach \citep{Bres:Clay:1993,Wolf:OCon:1993} is a
standard approach for fitting GLMMs, existing implementations such
as the glmmPQL() function in the MASS package in R do not include
MRF specifications in two dimensions and are not written to take advantage
of sparse precision matrices. In contrast the INLA methodology has
been implemented specifically for MRF models.

\subsubsection{Integrated Nested Laplace Approximation (INLA)}

\citep{Rue:etal:2009} present an approach to fitting GLMMs based
on nested Laplace approximations involving both the hyperparameters
and the latent process values. The result is estimation of the marginal
posterior densities of the random effects and the hyperparameters.
Note that this accounts for uncertainty in hyperparameters, in contrast
to the maximization done with the PQL approach. The INLA R package
(\url{www.r-inla.org}) can make use of sparsity in both $\bm{Q}$
and $\bm{K}$ and is therefore very computationally efficient. For
analyses using likelihood and prior models that are implemented in
INLA and for which one needs only marginal posteriors (or posteriors
of linear combinations), INLA is a promising option.

\subsubsection{MCMC}

While MCMC is a standard approach for Bayesian GLMMs, convergence
and mixing are often troublesome \citep{Chri:Waag:2002,Chri:etal:2006},
because of the high-dimensionality of the random effects, the dependence
between random effects (particularly when these represent spatial
or temporal structure), and cross-level dependence between random
effects and their hyperparameters ($\{\tau^{2},\kappa\}$ in this
work) \citep{Rue:Held:2005,Rue:etal:2009}. The sparse matrix calculations
possible with MRF models improve computational efficiency but do not
directly address convergence and mixing issues. 

\citet{Game:1997} describes the use of a weighted least squares proposal
for the fixed and random effects, also suggested in \citet[pp. 167-169]{Rue:Held:2005}
to deal with the dependence amongst the random effects, and \citet{Rue:Held:2005}
suggest combining this with a joint update of the hyperparameter and
the random effects to address the cross-level dependence. For high-dimensional
random effects vectors such as proposed here, using subblocks of $\bm{g}$
may also be helpful. To simulate draws of $\bm{g}$ from the posterior,
one might also use INLA to estimate the hyperparameters and then conditionally
draw samples of the fixed and random effects using MCMC.

One disadvantage of the TPS-MRF compared to the ICAR may be that the
additional smoothness makes it more difficult to accept MCMC proposals
for $\bm{g}$ because the value of the process in one grid cell is
more strongly constrained by the values in the other grid cells. The
Brownian motion-like behavior in the ICAR \citep{Besa:Mond:2005}
may help to decouple values of $\bm{g}$ for different grid cells,
similar to the strategy of adding a small amount of white noise to
the latent process \citep[e.g.,][]{Wikl:2002}, which \citet{Paci:2007b}
showed could improve MCMC mixing in a GP context.

\section{Examples}

\subsection{Point-level pollution modeling}

This example is based on the work of \citet{Paci:etal:2009}, who
modeled spatio-temporal variation in air pollution for the purpose
of predicting concentrations for use as the exposure values in a health
analysis. The data are average fine particulate matter over 2001-2002
at 339 monitoring stations in the northeast U.S., from the US EPA's
Air Quality System database. For this analysis, I averaged the 24
monthly average values and included only locations with at least 22
months of data. I compare the use of the MRF approach for modeling
point-level data (based on a 4 km resolution grid) with an additive
model built on a reduced rank thin plate spline (using the gam() function
from the mgcv package in R) and to universal kriging with both an
exponential covariance and a \matern covariance with $\nu=2$. For
the MRF models, I fitted a model of the form (\ref{eq:GLMM1}-\ref{eq:GLMM2})
with a linear link and normal likelihood (with independent, homoscedastic
errors) using the computational approach described in Section \ref{sub:Normal-data}.
I used the same set of covariates (log of the distance to nearest
road in two road size classes, percent urban land use in a local buffer,
log of elevation, and log of the estimated fine PM emissions within
a 10 km buffer) as in \citet{Paci:etal:2009}. I included the covariates
as linear terms for simplicity and because \citet{Paci:etal:2009}
found only a minor improvement when considering an additive nonparametric
structure for the covariates. I used ten-fold cross-validation to
assess hold-out error for 219 stations, with 120 stations in boundary
states forming a spatial buffer and always used in the training set,
to compare the models.

\begin{figure}
\includegraphics[scale=0.5]{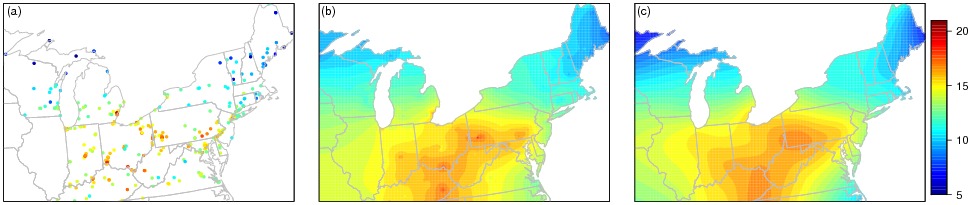}

\caption{PM observations (a) and fitted residual spatial surfaces (i.e., not
including the effect of covariates) for the ICAR (b) and TPS-MRF (c)
models.\label{fig:pmEx}}
\end{figure}

The models all gave very similar prediction results, with the square root of the mean squared
prediction errors being 1.30 for the gam() function, 1.34 and 1.29 for
universal kriging with the \matern ($\nu=2$) and exponential covariances
respectively, 1.32 for the TPS-MRF and 1.26 for the ICAR. These correspond
to cross-validation $R^{2}$ values of about 0.8 (0.787-0.811). The
slight apparent advantage for kriging with an exponential covariance
and for the ICAR model suggests the presence of fine-scale variability
that the other models smooth over. Fig.~\ref{fig:pmEx} shows the
estimates of residual spatial variability, not including the effect
of the spatially-varying covariates, illustrating that the ICAR model
fits local effects around the observations but also seems to generally
follow the large-scale pattern seen in the TPS-MRF. All the models
showed good hold-out prediction coverage, with the MRF models having
larger standard errors and therefore some overcoverage. 
The slight performance edge for the ICAR model here occurs in a setting in which it appears there is variation at multiple scales. In light of the scale dependence of the oracle and simulation results, which showed smaller values of $\rho$ and $\nu=0.5$
  tending to favor the ICAR over the TPS-MRF (Section~\ref{sub:Results-for-point}), this suggests the importance of the fine-scale variability to the predictive results here. The advantage for the ICAR may also relate to the fact that the observations are clustered (in metropolitan areas), a setting that the predictive results suggest works in favor of the ICAR model. 

\citet{Paci:Liu:2012} considered more complicated models of fine
PM over a variety of spatial and temporal domains using the MRF approach
outlined in this paper, including combining point-level monitoring
data with areal data from remote sensing and fitting a spatio-temporal
extension of the model proposed here.

\subsection{Area-level disease mapping}

This example is based on the work of \citet{Krie:etal:2006} and \citet{Hund:etal:2012},
who analyzed variation in breast cancer incidence in Los Angeles County,
California, US. Their analyses focused on the relationship between
census tract-level poverty and breast cancer incidence and changes
over time in that relationship, while accounting for residual spatial
variation. Here I modeled breast cancer incidence for data from 1998-2002
for white non-Hispanic women.

The data are counts of cancer incidence in the 5-year period. Following
\citet{Krie:etal:2006} and \citet{Hund:etal:2012}, I fitted Poisson
models with a log link, using the log of the expected counts as an
offset term,
\[
\log\mu_{i}=\log E_{i}+\beta_{0}+\bm{\beta}^{\top}\bm{\mbox{pov}}_{i}+\bm{K}_{i}^{\top}\bm{g},
\]
where $\mu_{i}$ and $E_{i}$ are the Poisson mean and the expected
number of cases in the $i$th census tract (CT). The expected numbers
were calculated based on internal age standardization, described in
\citet{Hund:etal:2012} and based on CT population (multiplied by
five, which assumes constant population over the 5-year period) from
the 2000 US Census. CT poverty was a five-level categorical variable
with $\bm{\mbox{pov}}_{i}$ being a vector of four indicator variables
determining the poverty category of the $i$th CT. The categorical
poverty variable is defined as follows: (1) $<5$\% of residents living
below the poverty line and more than 10\% of households having high
income (at least four times the US median household income), (2) $<5$\%
of residents living below the poverty line and less than 10\% of households
having high income, (3) 5.0-9.99\% of residents living below the poverty
line, (4) 10.0-19.99\% of residents living below the poverty line,
and (5) at least 20\% of residents living below the poverty line,
which was used as the baseline category. 

\begin{figure}
\includegraphics[scale=0.3]{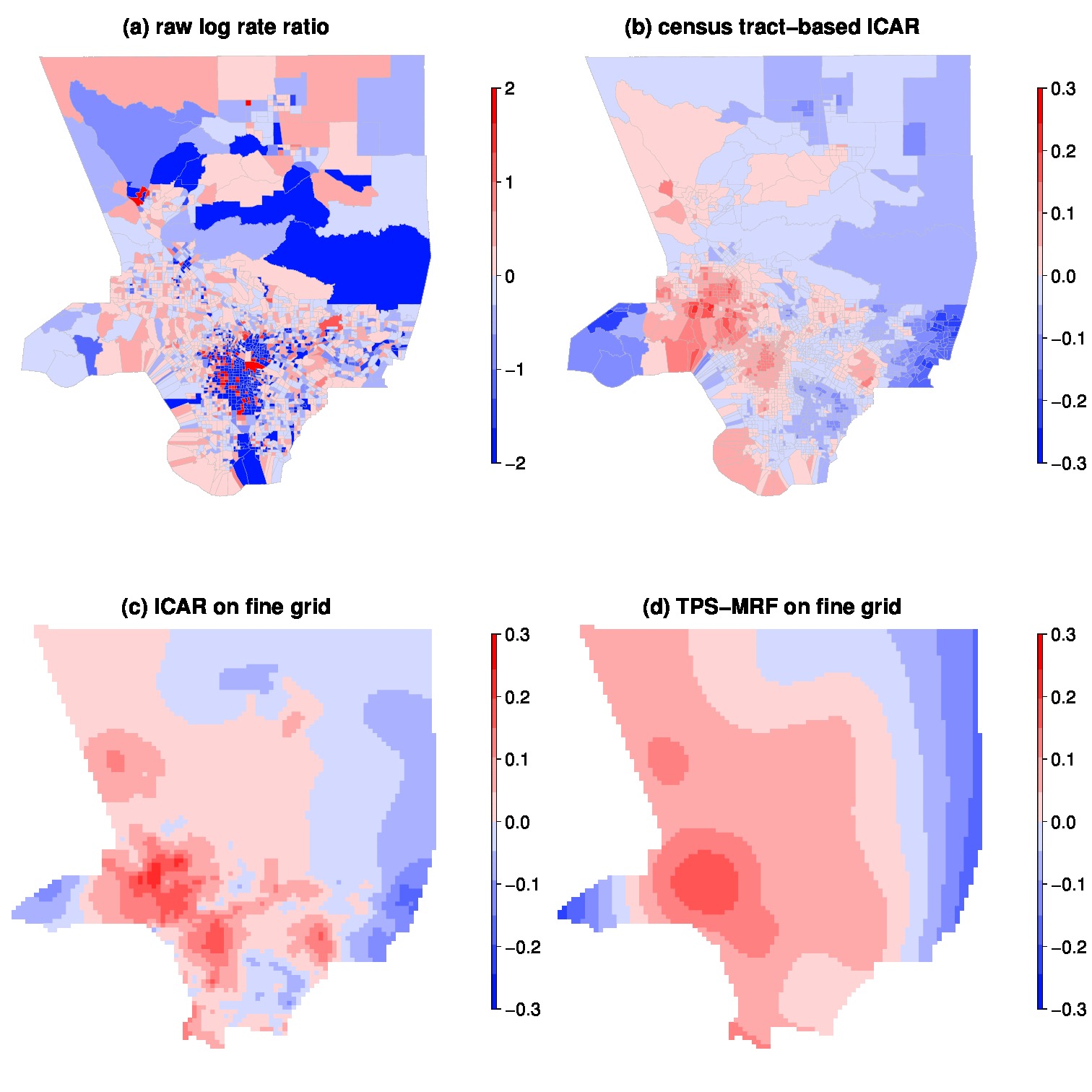}

\caption{(a) Raw log breast cancer incidence rate ratios by census tract in
Los Angeles County and estimated log incidence rate ratios for (b)
a standard ICAR model based on the census tract neighborhood structure,
(c) the ICAR model on a fine grid, and (d) the TPS-MRF model on a
fine grid. Note that the scale in (a) is different than the other
panels and that values with magnitude greater than two are censored.
Also note that the census tracts extend into the Pacific Ocean in
the southwestern portion of panels (a) and (b), while panels (c) and
(d) show only the land-based grid cells.\label{fig:laEx}}
\end{figure}

I used the INLA package in R to fit the ICAR and TPS-MRF models defined
on a fine grid with $99\times101$ cells of size $1.25^{2}$ $\mbox{km}^{2}$
and compared the results to a standard ICAR model based on the neighborhood
structure of the irregular census tracts, in all cases using the default
hyperparameter priors specified in INLA. The estimated coefficients
for the categorical poverty variable were very similar to those in
\citet{Hund:etal:2012}, with lower poverty CTs showing higher breast
cancer incidence. Fig.~\ref{fig:laEx} shows the log of the raw incidence
rate ratio (observed counts divided by expected number), compared
to the estimated log-incidence rate ratios, $\hat{\bm{g}}$, for the
three models. Values of zero indicate no departure from the expected
number based on the population and age distribution in the CT. The
census-tract-based ICAR and fine grid-based ICAR models show much
more spatial variability, while the TPS-MRF smooths quite a bit. DIC
and the summed log conditional predictive ordinate (CPO) values, $\sum_{i}\log P(Y_{i}|Y_{-i})$
(provided by the INLA package) suggest that the census-tract based
ICAR (DIC of 8438, logCPO of -4227) outperforms the grid-based ICAR
(8451, -4233) and TPS-MRF models (8476, -4241). The advantage of the
ICAR over the TPS-MRF is consistent with the simulation results for
areal data in which the ICAR performed better for the moderate value
of $\rho$ (Section~\ref{sub:Results-for-areal}).
Finally, using the \citet{Lind:etal:2011} approach to approximate
a \maternp-based GP on the same fine grid gave results in between
the grid-based MRF models, with a DIC of 8462 for $\nu=1$ (8467 for
$\nu=2$) and logCPO of -4237 (-4239 for $\nu=2$).

\section{Discussion}

I have presented a straightforward modeling approach for both areal
and point data that relates observations to an underlying smooth spatial
surface, represented as an MRF. One important result is that the analytic
comparison of various MRF structures indicates that higher-order neighborhood
structures that do not have the weighting structure of the TPS-MRF
do not produce smoother processes than the standard ICAR model. This
suggests that such models are not appealing for spatial modeling.
Given these results it would be useful to investigate the smoothing
properties of models that empirically choose neighborhood structures
\citep{Whit:Ghos:2009,Zhu:etal:2010}.

Based on the analytic and simulation assessment of predictive performance,
the TPS-MRF outperforms the ICAR model in many scenarios, in particular
with smoother surfaces (both in terms of the spatial range and differentiability),
as would be expected given that the TPS-MRF approximates a thin plate
spline. In both examples, however, the ICAR appeared to outperform
the TPS-MRF, perhaps because the true surface was relatively wiggly
in those contexts. One open question is how the models would perform
in a setting with variation at multiple resolutions; these results
suggest that the TPS-MRF better represents the large scale while the
ICAR model better captures fine-scale variation. How these balance
in terms of overall error would likely depend on the relative magnitude
of the variation at the different scales.

Spline models can do poorly in situations with large spatial gaps
(where large is relative to the spatial range of dependence in the
process being modeled) and on the boundary of the domain, as the estimation
of basis coefficients is poorly constrained by the data and influenced
by data at the extremes of the support of the basis functions. The
TPS-MRF model, by virtue of approximating a thin plate spline, can
have this unappealing behavior. In contrast a GP model, being a stationary
model, gives predictions that revert to the overall mean for prediction
locations far (relative to the estimated spatial range) from observations.
In many large datasets, for which the computational efficiency of
MRF models is appealing, including deterministic model output and
remote sensing observations, gaps are not present or tend to be small,
so the issue of extrapolation into large gaps may not be a concern.
Furthermore, my ad hoc experience suggests that gaps are less of a
problem in two dimensions than in one dimension, although the TPS-MRF
can have problems at the boundaries of the domain.

An appealing alternative to the MRF models presented here is the MRF
construction of \citet{Lind:etal:2011}, which approximates a Gaussian
process with \matern covariance for integer values of the \matern
smoothness (differentiability) parameter. I expect that much future
work with MRFs will involve this construction because of the added
flexibility of a representation that includes the GP range parameter.
However, I note that the simulations suggest that the TPS-MRF in some
cases outperforms an exact GP representation, and in the LA example,
I found that the ICAR models outperformed the \citet{Lind:etal:2011}
model. \citet{Lind:etal:2011} focus on a triangulation rather than
a rectangular grid, which has computational advantages when dealing
with irregular domains.

I have highlighted the advantages of using a smooth underlying surface
for areal data. These include the ability to deal with data aggregation
in a consistent manner and with spatial misalignment. Furthermore,
in many situations, the area boundaries are essentially arbitrary
relative to the process being measured, so the resulting neighborhood
structure is arbitrary as well. Rather it is appealing to imagine
an smooth underlying surface, with the areal units merely a measurement
artefact that is represented in the measurement model through the
mapping matrix, $\bm{K}$. In some cases, administrative units might
actually have a direct effect on the outcome, in which case more traditional
MRF models based on a single random effect per area and standard neighborhood
structures may be more appealing, although independent random effects
may be appealing in some cases.

Spatio-temporal modeling situations are of course very common. \citet{Paci:Liu:2012}
describe a spatio-temporal extension of the spatial models described
here that allows for autoregressive structure in time. In contrast
to many spatio-temporal models, the approach has a spatial mean that
is shared across time points, which \citet{Stei:Fang:1997} emphasize
is important for allowing one to properly characterize uncertainty
when aggregating over time periods.

\section*{Acknowledgments}

The author thanks Jeff Yanosky for the particulate matter dataset,
Jarvis Chen, Lauren Hund, and Brent Coull for access to the LA breast
cancer dataset, Steve Melly for GIS processing for the LA example,
and Finn Lindgren for helpful discussions. This work was funded by
NCI P01 Grant CA134294-01. 

\bibliographystyle{natbib}
\bibliography{draft4}

\end{document}